\documentclass{naturefig}
\usepackage{graphicx}

\newcommand{\msun}{M_{\odot}}
\newcommand{\lsun}{L_{\odot}}
\newcommand{\rsun}{R_{\odot}}
\newcommand{\krho}{k_{\rho}}
\newcommand{\tb}{T_{\rm b}}

\newcommand{\tch}{T_{\rm ch}}
\newcommand{\rct}{\tilde{R}}
\newcommand{\rch}{R_{\rm ch}}
\newcommand{\kt}{k_{\rm T}}
\newcommand{\etacr}{\eta_{\rm halt}}
\newcommand{\etaeq}{\eta_{\rm grav}}
\newcommand{\tf}{t_{\rm f}}
\newcommand{\tffc}{\overline{t}_{\rm ff}}
\newcommand{\mbar}{\overline{m}_*}
\newcommand{\sfrff}{\mbox{SFR}_{\rm ff}}
\newcommand{\avir}{\alpha_{\rm vir}}
\newcommand{\calm}{\mathcal{M}}
\newcommand{\cs}{c_s}
\newcommand{\tbone}{T_{\rm b,1}}
\newcommand{\mh}{m_{\rm H}}

\newcommand{\tkh}{t_{\rm KH}}

\newcommand{\kb}{k_{\rm B}}
\newcommand{\Lt}{\tilde{L}}
\newcommand{\ssb}{\sigma_{\rm SB}}
\newcommand{\Sigmath}{\Sigma_{\rm th}}
\newcommand{\psiimf}{\left\langle\psi\right\rangle_{\rm IMF}}
\newcommand{\nh}{n_{\rm H}}

\newcommand{\td}{T_{\rm d}}
\newcommand{\tg}{T_{\rm g}}
\newcommand{\sigmadz}{\sigma_{\rm d0}}
\newcommand{\Sigmathresh}{\Sigma_{\rm th}}
\newcommand{\ltsim}{\protect\raisebox{-0.5ex}{$\:\stackrel{\textstyle <}
	{\sim}\:$}}
\newcommand{\gtsim}{\protect\raisebox{-0.5ex}{$\:\stackrel{\textstyle >}
	{\sim}\:$}}

\newcommand{\apj}{\textit{Astrophys. J.}}
\newcommand{\apjl}{\textit{Astrophys. J. Lett.}}
\newcommand{\apjs}{\textit{Astrophys. J. Supp.}}
\newcommand{\mnras}{\textit{Mon. Not. R. Astron. Soc.}}

\newcommand{\aap}{\textit{Astron. \& Astronphys.}}

\newcommand{\nat}{\textit{Nature}}

\title{A Minimum Column Density of 1 g cm$^{-2}$ for Massive Star Formation}

\author{Mark R. Krumholz$^1$ \& Christopher F. McKee$^2$}

\begin{document}

\maketitle

\begin{affiliations}
\item Hubble Fellow; Astrophysics Department, Princeton University, Princeton, NJ 08544 and Astrophysics Department, UC Santa Cruz, Santa Cruz, CA 95064
\item Physics and Astronomy Departments, UC Berkeley, Berkeley, CA 94720
\end{affiliations}

\begin{abstract}
Massive stars are very rare, but their extreme luminosities make them both the only type of young star we can observe in distant galaxies and the dominant energy sources in the universe today. They form rarely because efficient radiative cooling keeps most star-forming gas clouds close to isothermal as they collapse, and this favors fragmentation into stars $\ltsim 1$ $\msun$ (solar mass; ref. \cite{larson05,jappsen05,bonnell06d}). Heating of a cloud by accreting low-mass stars within it can prevent fragmentation and allow formation of massive stars\cite{krumholz06b, krumholz07a}, but what properties a cloud must have to form massive stars, and thus where massive stars form in a galaxy, has not yet been determined. Here we show that only clouds with column densities $\gtsim 1$ g cm$^{-2}$ can avoid fragmentation and form massive stars. This threshold, and the environmental variation of the stellar initial mass function (IMF) that it implies, naturally explain the characteristic column densities of massive star clusters\cite{plume97,mueller02,shirley03,mckee03} and the difference between the radial profiles of H$\alpha$ and UV emission in galactic disks\cite{martin01, boissier06}. The existence of a threshold also implies that there should be detectable variations in the IMF with environment within the Galaxy and in the characteristic column densities of massive star clusters between galaxies, and that star formation rates in some galactic environments may have been systematically underestimated.
\end{abstract}

Consider a simple model system: a spherical gas cloud of mass $M$, column density $\Sigma$, radius $R=\sqrt{M/(\pi \Sigma)}$, and density profile $\rho\propto r^{-\krho}$, with a point source of luminosity $L$ at its center, representing the radiation output by stars beginning to form within it. In the limit $L\rightarrow 0$, the cloud falls to a background temperature $\tb$ set by the balance between cosmic ray heating and molecular and dust cooling. We are interested in the earliest stages of cloud collapse, so we adopt $\krho=1$. This puts most of the mass at low density, and is expected if clouds are in rough hydrostatic balance and obey the observed molecular cloud linewidth-size relation\cite{heyer04a} $\sigma\propto r^{q}$, where $\sigma$ is the velocity dispersion, $r$ is the size scale, and $q\approx 0.5$. However, any choice in the range $1 \leq \krho \leq 2$ yields the same qualitative conclusions.

The dust in a spherical cloud with a central source of illumination has a powerlaw temperature structure $T = \tch (r/\rch)^{-\kt}$, where $\tch$, $\rch$, and $\kt$ are functions of the cloud column density $\Sigma$, the light to mass ratio $\eta\equiv L/M$, and the dust opacity, which we characterize through a parameter $\delta$ that we define below\cite{chakrabarti05}. As we show in the Supplementary Information using a grain-gas energy exchange code\cite{neufeld95, young04, urban07}, at the high densities with which we are concerned, the gas temperature will be nearly identical to the dust temperature. The temperature will be everywhere greater than $\tb$ if
\begin{equation}
\label{etacr_eqn}
\tch(\eta,\Sigma,\delta) \left[\frac{R}{\rch(\eta,\Sigma,\delta)}\right]^{-\kt(\eta,\Sigma,\delta)} = \tb.
\end{equation}
Since $\kt$ is generally close to $0.4$ for strong sources of internal illumination and large $R/\rch$, a cloud satisfying this condition has an effective adiabatic index $\gamma\approx 1.4$ throughout its volume. Since even $\gamma \approx 1.1-1.2$ is sufficient to suppress fragmentation\cite{krumholz07a}, equation (\ref{etacr_eqn}) implicitly defines a critical light to mass ratio $\etacr$ above which fragmentation will halt in a cloud with a given $\Sigma$, $\delta$, and $\tb$. We describe our procedure for solving this equation in the Supplementary Information.

\begin{figure}
\begin{center}
\includegraphics[scale=1.5]{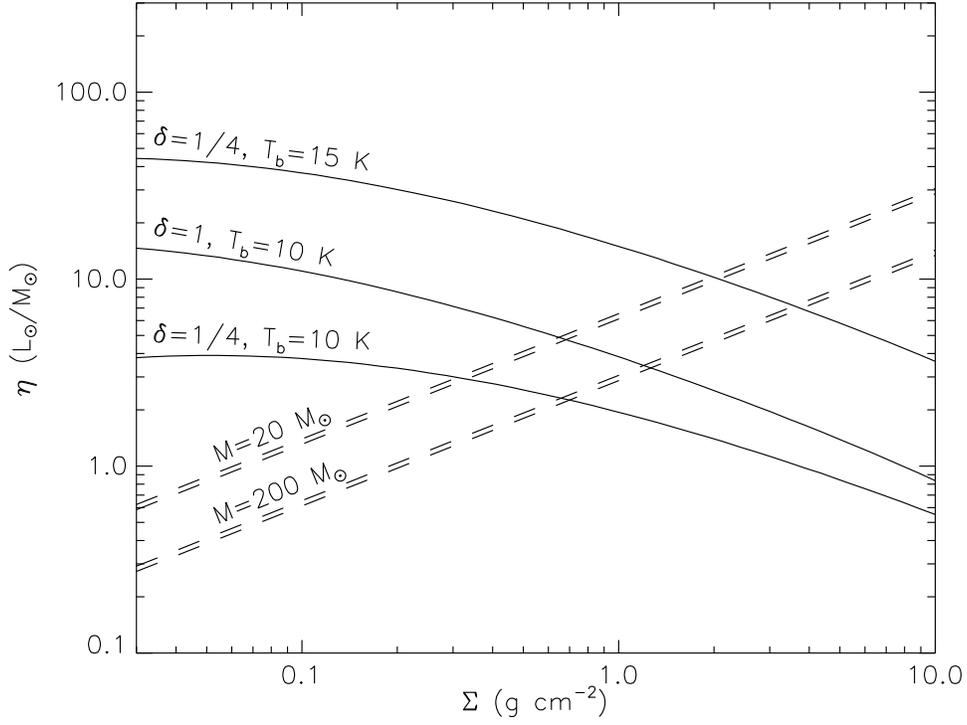}
\end{center}
\caption{
\label{etasigma}
Critical and equilibrium light to mass ratios versus cloud column density.
The plot shows the critical light to mass ratio $\etacr$ (\textit{solid lines}) and the equilibrium light to mass ratio $\etaeq$ due to low mass star formation (\textit{dashed lines}) as a function of the cloud column density $\Sigma$. The three curves for $\etacr$ are computed for $\delta=1$, $\tb=10$ K, for $\delta=1$, $\tb=15$ K, for $\delta=1/4$, $\tb=10$ K, and for $\delta=1/4$, $\tb=15$ K, as indicated. 
The two sets of curves for $\etaeq$ are computed for $M=20$ $\msun$ and $M=200$ $\msun$, as indicated, corresponding to the clouds that would be required to form $10$ $\msun$ and $100$ $\msun$ stars for a typical star-formation efficiency of 50\%.\cite{matzner00} In each pair the lower curve corresponds to $\tb=10$ K and the upper to $\tb=15$ K. Note that the background temperature can be higher than our assumed $\tb$ in regions near massive stars, but it is unclear whether massive stars in a cluster ever form close enough in time so that the first to form can affect the formation of subsequent ones. In the Orion Nebula\cite{huff06} and W3 Main\cite{feigelson07} clusters, all the stars larger than $10$ $\msun$ that remain today formed over a time spread $\ltsim 10^5$ yr. This is comparable to the formation time of a single massive star\cite{mckee02}, so that the last massive star to form must have been well-along by the time the first began to heat its envelope. Even in non-coeval clusters, our approach applies to the first massive stars.
}
\end{figure}

We approximate the infrared dust opacity as $\kappa=\delta \kappa_0 (\lambda_0/\lambda)^2$, where $\delta$ is a dimensionless number that we define to be unity at solar metallicity, $\lambda$ is the radiation wavelength, and $\lambda_0=100$ $\mu$m. Observations in the Milky Way indicate that, in cold regions where dust grains are coated with ice mantles, $\kappa_0\approx 0.54$ cm$^2$ g$^{-1}$ (ref. \cite{weingartner01, chakrabarti05}). Under Milky Way conditions the minimum temperature for interstellar gas is $\tb \approx 10$ K, with a weak density-dependence that we ignore for simplicity. In addition to the Milky Way case, we also consider $\delta=0.25$, $\tb=10$ K, appropriate for in a low metallicity galaxy today, and $\delta=0.25$, $\tb=15$ K, typical of a galaxy at $z\approx 6$ that has low metallicity but a temperature floor of 15 K imposed by the cosmic microwave background (CMB). Figure \ref{etasigma} shows the value of $\etacr$ computed for the three cases. We find that $\etacr$ declines with $\Sigma$ because at higher $\Sigma$ a cloud of fixed mass has a smaller radiating area and remains warmer at fixed luminosity.

\begin{figure}
\begin{center}
\includegraphics[scale=1.25]{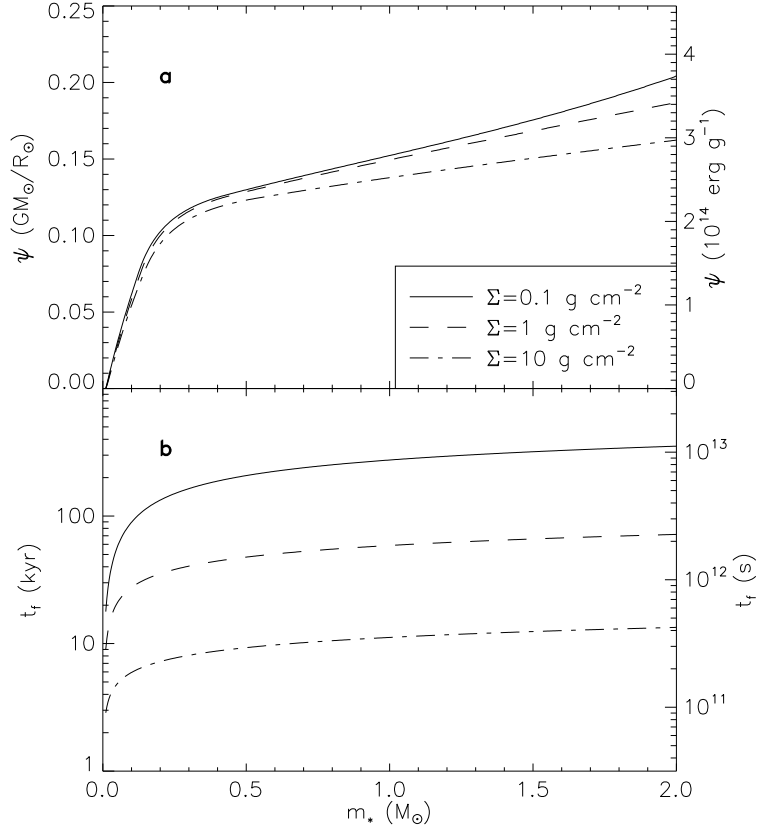}
\end{center}
\caption{
\label{startracks}
Energy per unit mass radiated and formation time versus protostellar mass.
The plot shows the energy per unit mass $\psi$ radiated away in the process of forming a star of mass $m_*$ (\textbf{a}), and the time required to form the star $\tf$ (\textbf{b}). The tracks shown are computed using a one-zone protostellar evolution code\cite{mckee03} applied to the accretion histories predicted by refs. \cite{mckee02,mckee03} using their fiducial parameters, for protostellar cores born in environments where the column density is $\Sigma=0.1$, $1.0$, or $10.0$ g cm$^{-2}$ as indicated. However, alternative accretion histories give qualitatively identical results. Note that $\psi$ is nearly independent of both accretion history and final stellar mass because the entropy distribution within a protostar and the protostellar mass-radius relation are nearly constant on time scales short compared to the Kelvin-Helmholtz time $\tkh$, and for low-mass stars $\tf \ll \tkh \approx 10$ Myr. This means that $\psi$, which is a measure of the gravitational energy released, is nearly independent of accretion history. Moreover, since low-mass stars have nearly linear mass-radius relations, $\psi$ is also nearly independent of the final stellar mass. Although our calculation of $\psi$ uses a code calibrated to solar metallicity, our results should also apply over a very wide range of metallicities because even for low metallicity stars $\tf\ll \tkh$.
}
\end{figure}

Clouds containing massive stars can have light-to-mass ratios of $100$ $\lsun/\msun$\cite{wu05}, more than sufficient to stop fragmentation, but we are interested in clouds where no massive stars have yet formed because fragmentation breaks all collapsing objects down to small masses. For a low-mass protostar the dominant energy source is gravitational potential energy radiated away by accreting gas. We plot the energy released per unit mass accreted, $\psi$, in Figure \ref{startracks}. Consider a cloud converting its mass into stars at a rate $\dot{M}_*$ with a mass distribution $dn/d\ln m_*$ and a mean mass $\mbar=\int m_* (dn/d\ln m_*)\, d\ln m_*$. Once the rate at which new stars in a cloud begin accreting balances the rate at which other stars reach their final mass and stop accreting, the light-to-mass ratio is
\begin{eqnarray}
\etaeq & = & \frac{1}{M} \left(\frac{\dot{M}_*}{\mbar}\right) \int \frac{dn}{d\ln m_*} \psi m_* \,d\ln m_* \\
& = & \frac{\sfrff}{\tffc} \psiimf,
\label{etaeq_eqn}
\end{eqnarray}
where $\sfrff=(\dot{M}_* \tffc)/M$ is the fraction of a cloud's mass that it turns into stars per mean density-free-fall time $\tffc$, and $\psiimf=\mbar^{-1} \int (dn/d\ln m_*) \psi m_* \, d\ln m_*$ is the IMF-averaged value of $\psi$. For a Chabrier IMF\cite{chabrier05} truncated at a maximum mass of $1$ $\msun$, $\psiimf = 2.1\times 10^{14}\mbox{ erg g}^{-1}=0.11 (G\msun/\rsun)$. Observations constrain $\sfrff$ to be a few percent\cite{tan06a, krumholz06c}, and in the Supplementary Information we use an analytic fitting formula\cite{krumholz05c} to estimate $\sfrff \approx 0.041 (M_2 \Sigma_0/\tbone^2)^{-0.08}$, where $M_2=M/(100\mbox{ }\msun)$, $\Sigma_0=\Sigma/(1\mbox{ g cm}^{-2})$, and $\tbone=\tb/(10\mbox{ K})$. Combining our estimates for $\psiimf$ and $\sfrff$ with the definition of the mean density free-fall time ($\tffc = \sqrt{3\pi/(32 G \overline{\rho})} = 38.6\,M_2^{1/4} \Sigma_0^{-3/4}$ kyr), we find that the light-to-mass ratio of a cloud powered by accretion onto low-mass stars is
\begin{equation}
\label{etaeq_eqn1}
\etaeq \approx 3.6\, M_2^{-0.33} \Sigma_0^{0.67} \tbone^{0.16} \left(\frac{\lsun}{\msun}\right).
\end{equation}
For the models shown in Figure \ref{startracks} a cloud reaches its equilibrium light-to-mass ratio $\ltsim 3 \tffc$ after star formation begins, and since $\sfrff \ltsim 0.05$ at most $\sim 15\%$ of the mass will have gone into low-mass stars at this point. If star formation accelerates in time, as predicted by some models\cite{bonnell04}, then $\etaeq$ will reach the value given by equation (\ref{etaeq_eqn1}) even earlier.

\begin{figure}
\begin{center}
\includegraphics[scale=1.5]{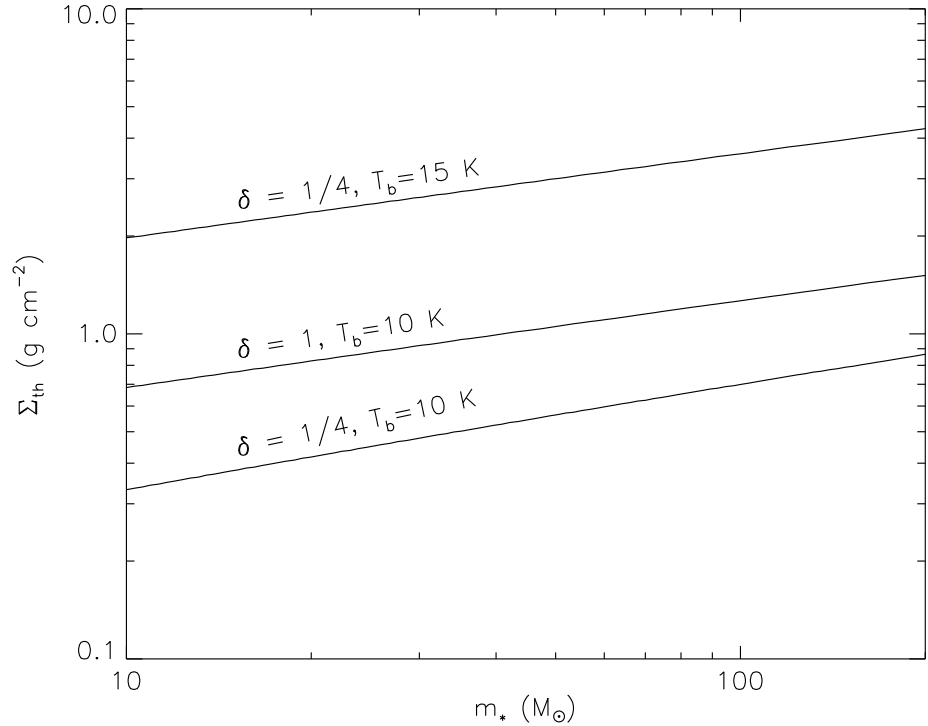}
\end{center}
\caption{
\label{msigma}
Threshold column density versus stellar mass.
The plot shows the threshold column density $\Sigmathresh$ required to form a star of mass $m_*$ for 
$\delta=1$, $\tb=10$ K, for $\delta=1/4$, $\tb=10$ K, and for $\delta=1/4$, $\tb=15$ K, as indicated. In making the plot we assumed an efficiency of 50\%\cite{matzner00}, so that the cloud mass required to make a star of mass $m_*$ is $M=2m_*$.
}
\end{figure}

In Figure \ref{etasigma}, we show $\etaeq$ computed for some typical parameters overplotted with $\etacr$. For each cloud mass $M$, we solve equations (\ref{etacr_eqn}) and (\ref{etaeq_eqn1}) to find the column density $\Sigmathresh$ such that $\etacr\geq\etaeq$, and plot the result in Figure \ref{msigma}. This is the threshold required for fragmentation to halt. We find that thresholds of $0.7-1.5$ g cm$^{-2}$ are required to form stars of $10-200$ $\msun$ under Milky Way conditions. Lower metallicity galaxies with comparable background temperatures require column densities that are a factor of $\sim 3$ smaller, while galaxies at $z\approx 6$ with low metallicity but high CMB temperatures require higher column densities by a similar factor. 

The existence of a threshold for massive star formation both explains current observations and predicts future ones. Regions with column densities $\gtsim 1$ g cm$^{-2}$ are rare even among star forming clouds, and contain only a small fraction of the molecular mass in the Galaxy, but our threshold explains why all nearby regions of massive star formation have $\Sigma$ at or above this value\cite{mckee03,plume97,mueller02,shirley03}. We further predict that clusters formed with $\Sigma \ll 1$ g cm$^{-2}$ should be deficient in massive stars. This is probably unobservable in individual low-column density clusters because they contain too few stars, but a statistical analysis of many clusters might reveal the effect.

Conversely, suppression of fragmentation should produce top-heavy IMFs at high gas column densities. This prediction can be tested by x-ray searches for low-mass protostars in high $\Sigma$ clouds that are not detected by Spitzer at 24 $\mu$m and therefore contain no massive protostars\cite{motte07}. We predict that any low-mass protostellar populations detected will constitute at most $15\%$ of the total mass. This prediction provides a sharp test for distinguishing our model from competitive accretion models, which predict that the mass of the most massive star forming in a cloud is related to the mass of low-mass stars around it by $(M_{\rm mass}/\msun) \approx (M_{\rm low-mass}/\msun)^{2/3}$ (ref. \cite{bonnell04}). Thus we would predict that a 100 $\msun$, $\Sigma > 1$ g cm$^{-2}$ cloud with no stars larger than $10$ $\msun$ should have a total stellar content below $15$ $\msun$, while competitive accretion would allow up to $42$ $\msun$ of low-mass stars. However, since radiation does not halt fragmentation until some low-mass stars have formed, we do expect most massive stars to form surrounded by a cluster.

Environmental variation at the top end of the IMF has even more profound consequences for extragalactic astronomy, since observations of distant galaxies are generally sensitive only to massive stars. The threshold explains why H$\alpha$ emission in galactic disks ends at sharp edges where the disks transition from gravitationally unstable to gravitational stable\cite{martin01}, but UV emission declines smoothly with radius and does not show a feature at the H$\alpha$ edge\cite{boissier06}. H$\alpha$ and UV are both tracers of recent star formation but are sensitive to different parts of the IMF. Outside the gravitational stability radius,  the molecular-to-atomic surface density ratio drops sharply\cite{braine07}, probably because purely local instabilities create small molecular clouds but not giant complexes like those in disks' inner parts. Since compressing gas to column densities of $\Sigmathresh$ requires a huge amount of weight that can only be provided by such giant complexes, their absence will selectively suppress the formation of the most massive stars, leading to a truncated IMF. If in such a region no stars larger than, for example, $15$ $\msun$ were to form, this would reduce UV emission by $\sim 50\%$ but would eliminate more than $99\%$ of the H$\alpha$ light\cite{parravano03}, explaining the sharp drop in H$\alpha$ but not UV.

An important corollary to this point is that estimates of the star formation rate assuming a standard IMF in regions that do not contain giant clouds, such as much of the volume of dwarf galaxies and the outer parts of disk galaxies, may be systematically too low. At this point our theory is too approximate to allow a precise calculation of the extent of the underestimate.

Our calculation of $\Sigmathresh$ also enables us to predict how the characteristic column densities of young clusters containing massive stars will vary with Galactic environment. We predict that clusters that have cleared their gas but not yet dynamically expanded should show a minimum column density near $\Sigmathresh$ (probably slightly below $\Sigmathresh$ due to gas removal), and this should be lower at low metallicity and higher at high redshift as indicated in Figure \ref{msigma}.


\begin{addendum}
\item[Supplementary Information]is linked
to the online version of the paper at \\www.nature.com/nature.
\item[Acknowledgements]We acknowledge S. Boissier, I. Bonnell, B. Elmegreen, E. Feigelson, and C. Martin for discussions. We thank A. Urban, N. Evans, and S. Doty for providing a copy of their grain-gas coupling code. This work was supported by NASA through the Hubble Fellowship program and by the NSF. Parts of this work were performed while the authors were in residence at the Kavli Institute for Theoretical Physics at UCSB.
\item[Author Information]Reprints and Permissions information
is available at\\npg.nature.com/reprintsandpermissions. Authors declare
they have no competing financial interests. Correspondence and
requests for materials should be addressed to
M.R.K. \\(krumholz@astro.princeton.edu).
\end{addendum}

\clearpage

\spacing{1}

\clearpage

\clearpage

\clearpage

\spacing{2}
\setcounter{equation}{0}

{\Large\bfseries\noindent\sloppy \textsf{Supplementary Discussion}}

In this Supplementary Discussion, we give more detailed derivations of several of the results in our letter.

First, we describe in detail our method of calculating the temperature structure of our clouds and determining when they are dominated by internal illumination. As discussed in the main text, we consider a cloud of mass $M$, column density $\Sigma$, radius $R=\sqrt{M/(\pi \Sigma)}$, and density profile $\rho\propto r^{-\krho}$. The gas in the cloud is mixed with dust, which provides an opacity per unit mass $\kappa=\delta \kappa_0 (\lambda_0/\lambda)^{\beta}$, where $\lambda$ is the radiation wavelength, $\lambda_0=100$ $\mu$m, $\kappa_0 = 0.54$ cm$^{2}$ g$^{-1}$, $\beta=2$, and the choice $\delta=1$ corresponds to typical properties in cold gas in the Milky Way, where ice mantles on dust grains double the opacity compared to that in the diffuse interstellar medium\cite{weingartner01}. For convenience we define $T_0=hc/(\kb\lambda_0)=144$ K. There is a source of luminosity $L$ located at the cloud center, and we define $\eta=L/M$.

To compute the cloud temperature, we first estimate the dust temperature using the method of ref. \cite{chakrabarti05}, who show that the dust temperature profile in a cloud of this sort takes an approximately powerlaw form:
\begin{equation}
T = \tch \left(\frac{r}{\rch}\right)^{-\kt}.
\end{equation}
The characteristic emission radius $\rch$ and emission temperature $\tch$ are defined by two conditions: the total luminosity must be $L=4\pi \Lt \rch^2 \ssb \tch^4$, where $\Lt$ is a constant of order unity to be determined, and the optical depth from the cloud edge to $\rch$ must be unity for radiation whose wavelength is $\lambda=hc/(\kb \tch)$. Together these conditions imply 
\begin{eqnarray}
\label{rcteqn}
\rct \equiv\frac{R}{\rch} & = & \left\{
\left(\frac{\eta}{4\ssb\Lt}\right)^{\beta} \Sigma^{4+\beta} \left[\frac{(3-\krho)\kappa_0}{4(\krho-1) T_0^{\beta}} \right]^4\right\}^{-\alpha} \\
\label{tcheqn}
\tch & = & \left\{\left(\frac{\eta}{4\ssb\Lt}\right)^{\krho-1} \Sigma^{\krho-3}
\left[\frac{4(\krho-1) T_0^{\beta}}{(3-\krho)\kappa_0}\right]^2\right\}^{\alpha},
\end{eqnarray}
where
$\alpha \equiv 1/[2\beta+4(\krho-1)]$. The approximation
\begin{equation}
\label{lteqn}
\Lt \approx 1.6 \rct^{0.1}
\end{equation}
reproduces numerical solutions of the transfer equation very accurately, so we adopt it. The temperature structure within the cloud takes an approximately powerlaw form, $T = \tch (r/R)^{-\kt}$, with $\kt$ well-fit by the approximation
\begin{equation}
\label{kteqn}
\kt \approx \frac{0.48 \krho^{0.05}}{\rct^{0.02\krho^{1.09}}} + \frac{0.1\krho^{5.5}}{\rct^{0.7 \krho^{1.9}}}.
\end{equation}
Note that the approximations in ref. \cite{chakrabarti05} break down and the equations become singular as $\krho\rightarrow 1$, but numerical solutions of the transfer equation show that the temperature profiles for $\krho=1$ and $\krho=1.1$ are nearly identical. Thus, we handle the case $\krho < 1.1$ by approximating it with $\krho=1.1$. Also note that equation (\ref{lteqn}) is a slightly different approximation than that given in ref.\ \cite{chakrabarti05}. This approximation provides a somewhat better fit than the approximation given there (S.\ Chakrabarti, private communication, 2007). 

If the dust and gas temperatures are equal, it is straightforward to solve equation (1) in the main text, so we proceed by assuming that they are equal and then check that assumption. Equations (\ref{rcteqn}) - (\ref{lteqn}) constitute three equations in the unknowns $\rch$, $\tch$, and $\Lt$, which we may solve for specified values of $\Sigma$, $\eta$, and $\krho$. Together with equation (\ref{kteqn}), this fully specifies the dust temperature profile in the cloud. To solve equation (1) in the main text for $\etacr$, we simply fix $\delta$ and $\Sigma$, and iterate to obtain the value of $\eta$ that satisfies the equation.

Our second calculation in the Supplementary Discussion is a check of the assumption that the dust and gas temperatures are equal using both a simple analytic estimate and using a detailed numerical calculation of dust-gas energy exchange. We perform this check for the threshold column densities $\Sigmath$ that we obtain under the assumption that the dust and gas temperatures are equal. In the vicinity of an illuminating source that heats the dust, the gas temperature will be determined by a competition between the dominant heating process, collisions with warm dust grains, and the dominant cooling process, molecular cooling. The former will be least effective and the latter most effective at the low density edge of a cloud, so we focus our attention at cloud edges, where the temperatures are near $\tb$ and the densities are lowest. The volumetric gas heating rate by dust collisions is approximately\cite{young04}
\begin{equation}
\label{graingas}
\Gamma_{\rm gd} = 9.0\times 10^{-34} \nh^2 \tg^{0.5}
\sigmadz 
\left[1 - 0.8\exp\left(-\frac{75\mbox{ K}}{\tg}\right)\right] 
(\td - \tg)
\mbox{ erg cm}^{-3}\mbox{ s}^{-1},
\end{equation}
where $\nh$ is the number density of hydrogen nuclei, $\tg$ is the gas temperature, $\td$ is the dust temperature, and $\sigmadz$ is the average dust cross section per baryon, normalized to the fiducial Milky Way value, $6.09\times 10^{-22}\mbox{ cm}^{-2}$. We make the simple assumption that the dust-grain cross section varies with metallicity in the same way as the total dust opacity, so we adopt $\sigmadz = \delta$.

The dominant dust cooling process is molecular line emission, with CO dominating at lower densities and optical depths and other species taking over at higher densities. The exact cooling rate is a very complex function of density, temperature, and optical depth, but numerical radiative transfer and molecular excitation calculations\cite{neufeld95} show that, for a gas temperature near $\tg=10$ K, the total cooling rate has a roughly constant value of
\begin{equation}
\Lambda_{\rm mol} \approx 10^{-27} \delta\nh\mbox{ erg cm}^{-3}\mbox{ s}^{-1}
\end{equation}
for $\nh$ in the range $10^{3}-10^{7}$ cm$^{-3}$. We have again taken the molecular cooling rate to be simply proportional to the metallicity. Adopting this cooling rate, and solving for the gas energy balance by setting $\Gamma_{\rm gd}=\Lambda_{\rm mol}$, we find
\begin{equation}
\td-\tg \approx \frac{3.5\times 10^5}{\nh}\mbox{ K}.
\end{equation}
Thus, the gas and dust temperature will be identical to within 1 K for densities $\nh \gtsim 3.5\times 10^5$ cm$^{-3}$, roughly independent of $\delta$.

We can now compare this to the minimum densities found in gas clouds at the critical column density. For a cloud of mass $M$ and column density $\Sigma$, the number density at the cloud edge is
\begin{equation}
\nh = \frac{3-\krho}{4\mu} \sqrt{\frac{\pi \Sigma^3}{M}},
\end{equation}
where $\mu=2.34\times 10^{-24}$ g is the mean mass per H nucleus for a gas of the standard cosmic abundance. For stars in the mass range $10-200$ $\msun$, for the threshold column densities $\Sigmath$ shown in Figure 3 of the main text, the minimum values of $\nh$ are $7.9\times 10^5$, $3.4 \times 10^5$, and $3.8\times 10^6$ cm$^{-3}$ for the cases $\delta=1$ and $\tb=10$ K, $\delta=0.25$ and $\tb=10$ K, and $\delta=0.25$ and $\tb=15$ K, respectively. These values suggest that the cloud volume densities are high enough for our threshold cases that our assumption of dust-gas temperature equality is well-justified, although the case $\delta=0.25$, $\tb=10$ K is perhaps marginal.

We further check this approximation by computing the gas temperature using a detailed gas-temperature calculation code\cite{young04,urban07}. The code takes as input fixed spherically-symmetric density and dust temperature profiles, and computes the resulting gas temperature profile. It includes grain-gas energy exchange, cosmic ray heating, and line cooling from a large number of species, including an approximate treatment of radiative trapping effects in the Sobolev approximation. In the code, we set the cosmic ray ionization rate to its Milky Way value, we set the grain-gas energy exchange rate to the value given by equation (\ref{graingas}), and we scale the tabulated molecular cooling rate by $\delta$ under the assumption that the molecular cooling rate is simply proportional to metallicity. We also assume that our clouds are embedded in an environment where the column density is $10\%$ of $\Sigmath$.

Supplementary Figure 1 shows the cloud temperature profiles and the fractional difference between the dust and gas temperatures for the cases $\delta=1$ and $\tb=10$ K, $\delta=0.25$ and $\tb=10$ K, and $\delta=0.25$ and $\tb=15$ K. The cases shown are for clouds with mass $M=200$ $\msun$, which we approximate will produce $100$ $\msun$ stars if they do not fragment\cite{matzner00}, but different cloud masses do not produce qualitatively different results. As the plots show, the dust and gas temperatures agree to within a few percent even at the cloud edges. Thus, our assumption of dust-gas temperature equality is well-founded. In fact, the detailed calculation shows that our analytic approximation overestimates the grain-gas temperature difference. This is likely because in our analytic estimate we neglected cosmic ray heating, which becomes comparable to the assumed cooling rate $\Lambda_{\rm mol}$ when the temperature is near $\tb$, and because our analytic estimate of $\Lambda_{\rm mol}$ is based on calculations for somewhat lower column density regions than the ones we are considering, and the cooling rate is higher at lower column density due to decreased line opacity.

The third and final calculation we present in the Supplementary Discussion is a derivation of the estimated total rate of star formation in our centrally condensed cloud. We base our estimate on a fit to the star formation rate in simulations of driven turbulent motion in periodic boxes\cite{krumholz05c}
\begin{equation}
\label{sfrffuniform}
\mbox{SFR}_{\rm ff-uniform} \approx 0.073 \avir^{-0.68} \calm^{-0.32},
\end{equation}
where $\avir$ is the cloud virial ratio and $\calm$ is its one-dimensional Mach number on scales comparable to the size of the entire cloud. The definition of the virial ratio is
\begin{equation}
\avir=\frac{5 \calm^2 \cs^2 R}{G M},
\end{equation}
where $\cs$ is the cloud's isothermal sound speed, which we compute at $\tb$. Following observations, we adopt $\avir=1.3$ as typical of star-forming clouds\cite{mckee03}, and this implies that
\begin{equation}
\calm = \frac{1}{\cs} \left(\frac{\pi \avir^2 G^2 M \Sigma}{25}\right)^{1/4} = 6.2 \left(\frac{M_2 \Sigma_0}{\tbone^2}\right)^{1/4},
\end{equation}
where $M_2=M/(100\mbox{ }\msun)$, $\Sigma_0=\Sigma/(1\mbox{ g cm}^{-2})$, and $\tbone=\tb/(10\mbox{ K})$. In the numerical evaluation, we have taken the mean mass per particle of the gas to be $2.33 \mh$, appropriate for a gas of the standard cosmic abundance. Using this value of $\calm$ in equation (\ref{sfrffuniform}) gives an estimate of $\sfrff$ on the outer scale of the cloud:
\begin{equation}
\mbox{SFR}_{\rm ff-out} \approx 0.034 \left(\frac{M_2 \Sigma_0}{\tbone^2}\right)^{-0.08}.
\end{equation}
This would be the value of $\sfrff$ if the cloud were uniform. For a centrally-concentrated cloud, if one assumes that equation (\ref{sfrffuniform}) applies locally everywhere within the cloud, the star formation rate is enhanced over $\mbox{SFR}_{\rm ff-out}$ by a factor of $(3-\krho)^{3/2} / [2.3(2-\krho)]$ (ref. \cite{tan06a}). This enhancement occurs because the mass-averaged free-fall time is higher and the mass-averaged Mach number is lower for a centrally-concentrated cloud than for a uniform one. Applying this factor to $\mbox{SFR}_{\rm ff-uniform}$ for $\krho=1$, we arrive at our final estimate for $\sfrff$ for the cloud:
\begin{equation}
\sfrff \approx 0.041 \left(\frac{M_2 \Sigma_0}{\tbone^2}\right)^{-0.08}.
\end{equation}

\clearpage

\renewcommand{\figurename}{Supplementary Figure}
\setcounter{figure}{0}
\spacing{1}

\begin{figure}
\begin{center}
\includegraphics[scale=1.25]{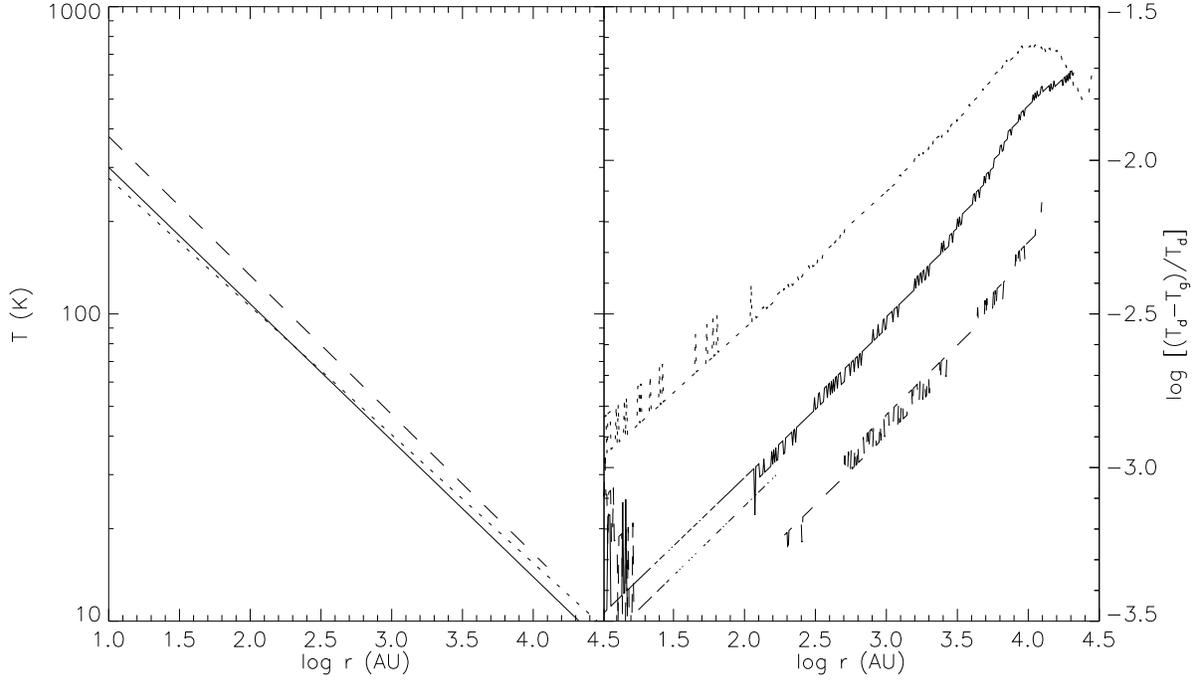}
\end{center}
\caption{
\label{dustgasprof}
Dust temperature and dust-gas temperature difference in a gas cloud. The left panel shows the dust temperature as a function of radius computed for clouds of mass $M=200$ $\msun$ with critical column densities, for the cases $\delta=1$ and $\tb=10$ K (\textit{solid lines}), $\delta=0.25$ and $\tb=10$ K (\textit{dotted lines}), and $\delta=0.25$ and $\tb=15$ K (\textit{dashed lines}). We do not show the gas temperature profile separately because the curves so closely overlay one another as to be indistinguishable. In the right panel, we show the fractional difference in dust and gas temperatures $(\td-\tg)/\td$ as a function of radius.
}
\end{figure}


\begin{thebibliography}{10}
\expandafter\ifx\csname url\endcsname\relax
  \def\url#1{\texttt{#1}}\fi
\expandafter\ifx\csname urlprefix\endcsname\relax\def\urlprefix{URL }\fi
\providecommand{\bibinfo}[2]{#2}
\providecommand{\eprint}[2][]{\url{#2}}

\bibitem{larson05}
\bibinfo{author}{{Larson}, R.~B.}
\newblock \bibinfo{title}{{Thermal physics, cloud geometry and the stellar
  initial mass function}}.
\newblock \emph{\bibinfo{journal}{\mnras}} \textbf{\bibinfo{volume}{359}},
  \bibinfo{pages}{211--222} (\bibinfo{year}{2005}).

\bibitem{jappsen05}
\bibinfo{author}{{Jappsen}, A.-K.}, \bibinfo{author}{{Klessen}, R.~S.},
  \bibinfo{author}{{Larson}, R.~B.}, \bibinfo{author}{{Li}, Y.} \&
  \bibinfo{author}{{Mac Low}, M.-M.}
\newblock \bibinfo{title}{{The stellar mass spectrum from non-isothermal
  gravoturbulent fragmentation}}.
\newblock \emph{\bibinfo{journal}{\aap}} \textbf{\bibinfo{volume}{435}},
  \bibinfo{pages}{611--623} (\bibinfo{year}{2005}).

\bibitem{bonnell06d}
\bibinfo{author}{{Bonnell}, I.~A.}, \bibinfo{author}{{Clarke}, C.~J.} \&
  \bibinfo{author}{{Bate}, M.~R.}
\newblock \bibinfo{title}{{The Jeans mass and the origin of the knee in the
  IMF}}.
\newblock \emph{\bibinfo{journal}{\mnras}} \textbf{\bibinfo{volume}{368}},
  \bibinfo{pages}{1296--1300} (\bibinfo{year}{2006}).

\bibitem{krumholz06b}
\bibinfo{author}{{Krumholz}, M.~R.}
\newblock \bibinfo{title}{{Radiation Feedback and Fragmentation in Massive
  Protostellar Cores}}.
\newblock \emph{\bibinfo{journal}{\apjl}} \textbf{\bibinfo{volume}{641}},
  \bibinfo{pages}{L45--L48} (\bibinfo{year}{2006}).

\bibitem{krumholz07a}
\bibinfo{author}{{Krumholz}, M.~R.}, \bibinfo{author}{{Klein}, R.~I.} \&
  \bibinfo{author}{{McKee}, C.~F.}
\newblock \bibinfo{title}{{Radiation-Hydrodynamic Simulations of Collapse and
  Fragmentation in Massive Protostellar Cores}}.
\newblock \emph{\bibinfo{journal}{\apj}} \textbf{\bibinfo{volume}{656}},
  \bibinfo{pages}{959--979} (\bibinfo{year}{2007}).

\bibitem{plume97}
\bibinfo{author}{{Plume}, R.}, \bibinfo{author}{{Jaffe}, D.~T.},
  \bibinfo{author}{{Evans}, N.~J.}, \bibinfo{author}{{Martin-Pintado}, J.} \&
  \bibinfo{author}{{Gomez-Gonzalez}, J.}
\newblock \bibinfo{title}{{Dense Gas and Star Formation: Characteristics of
  Cloud Cores Associated with Water Masers}}.
\newblock \emph{\bibinfo{journal}{\apj}} \textbf{\bibinfo{volume}{476}},
  \bibinfo{pages}{730--749} (\bibinfo{year}{1997}).

\bibitem{mueller02}
\bibinfo{author}{{Mueller}, K.~E.}, \bibinfo{author}{{Shirley}, Y.~L.},
  \bibinfo{author}{{Evans}, N.~J.} \& \bibinfo{author}{{Jacobson}, H.~R.}
\newblock \bibinfo{title}{{The Physical Conditions for Massive Star Formation:
  Dust Continuum Maps and Modeling}}.
\newblock \emph{\bibinfo{journal}{\apjs}} \textbf{\bibinfo{volume}{143}},
  \bibinfo{pages}{469--497} (\bibinfo{year}{2002}).

\bibitem{shirley03}
\bibinfo{author}{{Shirley}, Y.~L.}, \bibinfo{author}{{Evans}, N.~J.},
  \bibinfo{author}{{Young}, K.~E.}, \bibinfo{author}{{Knez}, C.} \&
  \bibinfo{author}{{Jaffe}, D.~T.}
\newblock \bibinfo{title}{{A CS J=$5\rightarrow 4$ Mapping Survey Toward
  High-Mass Star-forming Cores Associated with Water Masers}}.
\newblock \emph{\bibinfo{journal}{\apjs}} \textbf{\bibinfo{volume}{149}},
  \bibinfo{pages}{375--403} (\bibinfo{year}{2003}).

\bibitem{mckee03}
\bibinfo{author}{{McKee}, C.~F.} \& \bibinfo{author}{{Tan}, J.~C.}
\newblock \bibinfo{title}{{The Formation of Massive Stars from Turbulent
  Cores}}.
\newblock \emph{\bibinfo{journal}{\apj}} \textbf{\bibinfo{volume}{585}},
  \bibinfo{pages}{850--871} (\bibinfo{year}{2003}).

\bibitem{martin01}
\bibinfo{author}{{Martin}, C.~L.} \& \bibinfo{author}{{Kennicutt}, R.~C.}
\newblock \bibinfo{title}{{Star Formation Thresholds in Galactic Disks}}.
\newblock \emph{\bibinfo{journal}{\apj}} \textbf{\bibinfo{volume}{555}},
  \bibinfo{pages}{301--321} (\bibinfo{year}{2001}).

\bibitem{boissier06}
\bibinfo{author}{{Boissier}, S.} \emph{et~al.}
\newblock \bibinfo{title}{{Radial variation of attenuation and star formation
  in the largest late-type disks observed with GALEX}}.
\newblock \emph{\bibinfo{journal}{\apjs}}  (\bibinfo{year}{2006}).
\newblock \bibinfo{note}{In press, astro-ph/0609071}.

\bibitem{heyer04a}
\bibinfo{author}{{Heyer}, M.~H.} \& \bibinfo{author}{{Brunt}, C.~M.}
\newblock \bibinfo{title}{{The Universality of Turbulence in Galactic Molecular
  Clouds}}.
\newblock \emph{\bibinfo{journal}{\apjl}} \textbf{\bibinfo{volume}{615}},
  \bibinfo{pages}{L45--L48} (\bibinfo{year}{2004}).

\bibitem{chakrabarti05}
\bibinfo{author}{{Chakrabarti}, S.} \& \bibinfo{author}{{McKee}, C.~F.}
\newblock \bibinfo{title}{{Far-Infrared SEDs of Embedded Protostars and Dusty
  Galaxies. I. Theory for Spherical Sources}}.
\newblock \emph{\bibinfo{journal}{\apj}} \textbf{\bibinfo{volume}{631}},
  \bibinfo{pages}{792--808} (\bibinfo{year}{2005}).

\bibitem{neufeld95}
\bibinfo{author}{{Neufeld}, D.~A.}, \bibinfo{author}{{Lepp}, S.} \&
  \bibinfo{author}{{Melnick}, G.~J.}
\newblock \bibinfo{title}{{Thermal Balance in Dense Molecular Clouds: Radiative
  Cooling Rates and Emission-Line Luminosities}}.
\newblock \emph{\bibinfo{journal}{\apjs}} \textbf{\bibinfo{volume}{100}},
  \bibinfo{pages}{132} (\bibinfo{year}{1995}).

\bibitem{young04}
\bibinfo{author}{{Young}, K.~E.}, \bibinfo{author}{{Lee}, J.-E.},
  \bibinfo{author}{{Evans}, N.~J., II}, \bibinfo{author}{{Goldsmith}, P.~F.} \&
  \bibinfo{author}{{Doty}, S.~D.}
\newblock \bibinfo{title}{{Probing Pre-Protostellar Cores with Formaldehyde}}.
\newblock \emph{\bibinfo{journal}{\apj}} \textbf{\bibinfo{volume}{614}},
  \bibinfo{pages}{252--266} (\bibinfo{year}{2004}).

\bibitem{urban07}
\bibinfo{author}{{Urban}, A.}, \bibinfo{author}{{Evans}, N.~J., II} \&
  \bibinfo{author}{{Doty}, S.~D.}
\newblock \bibinfo{title}{{A Parameter Study of the Dust and Gas Temperature in
  a Field of Young Stars}}.
\newblock \emph{\bibinfo{journal}{\apj}}  (\bibinfo{year}{2007}).
\newblock \bibinfo{note}{Submitted, arXiv:0710.3906}.

\bibitem{weingartner01}
\bibinfo{author}{{Weingartner}, J.~C.} \& \bibinfo{author}{{Draine}, B.~T.}
\newblock \bibinfo{title}{{Dust Grain-Size Distributions and Extinction in the
  Milky Way, Large Magellanic Cloud, and Small Magellanic Cloud}}.
\newblock \emph{\bibinfo{journal}{\apj}} \textbf{\bibinfo{volume}{548}},
  \bibinfo{pages}{296--309} (\bibinfo{year}{2001}).

\bibitem{wu05}
\bibinfo{author}{{Wu}, J.} \emph{et~al.}
\newblock \bibinfo{title}{{Connecting Dense Gas Tracers of Star Formation in
  our Galaxy to High-z Star Formation}}.
\newblock \emph{\bibinfo{journal}{\apjl}} \textbf{\bibinfo{volume}{635}},
  \bibinfo{pages}{L173--L176} (\bibinfo{year}{2005}).

\bibitem{chabrier05}
\bibinfo{author}{{Chabrier}, G.}
\newblock \bibinfo{title}{{The Initial Mass Function: from Salpeter 1955 to
  2005}}.
\newblock In \bibinfo{editor}{{Corbelli}, E.}, \bibinfo{editor}{{Palla}, F.} \&
  \bibinfo{editor}{{Zinnecker}, H.} (eds.) \emph{\bibinfo{booktitle}{The
  Initial Mass Function 50 Years Later}}, vol. \bibinfo{volume}{327} of
  \emph{\bibinfo{series}{Astrophysics and Space Science Library}},
  \bibinfo{pages}{41} (\bibinfo{year}{2005}).

\bibitem{tan06a}
\bibinfo{author}{{Tan}, J.~C.}, \bibinfo{author}{{Krumholz}, M.~R.} \&
  \bibinfo{author}{{McKee}, C.~F.}
\newblock \bibinfo{title}{{Equilibrium Star Cluster Formation}}.
\newblock \emph{\bibinfo{journal}{\apjl}} \textbf{\bibinfo{volume}{641}},
  \bibinfo{pages}{L121--L124} (\bibinfo{year}{2006}).

\bibitem{krumholz06c}
\bibinfo{author}{{Krumholz}, M.~R.} \& \bibinfo{author}{{Tan}, J.~C.}
\newblock \bibinfo{title}{{Slow Star Formation in Dense Gas: Evidence and
  Implications}}.
\newblock \emph{\bibinfo{journal}{\apj}} \textbf{\bibinfo{volume}{654}},
  \bibinfo{pages}{304--315} (\bibinfo{year}{2007}).

\bibitem{krumholz05c}
\bibinfo{author}{{Krumholz}, M.~R.} \& \bibinfo{author}{{McKee}, C.~F.}
\newblock \bibinfo{title}{{A General Theory of Turbulence-regulated Star
  Formation, from Spirals to Ultraluminous Infrared Galaxies}}.
\newblock \emph{\bibinfo{journal}{\apj}} \textbf{\bibinfo{volume}{630}},
  \bibinfo{pages}{250--268} (\bibinfo{year}{2005}).

\bibitem{bonnell04}
\bibinfo{author}{{Bonnell}, I.~A.}, \bibinfo{author}{{Vine}, S.~G.} \&
  \bibinfo{author}{{Bate}, M.~R.}
\newblock \bibinfo{title}{{Massive star formation: nurture, not nature}}.
\newblock \emph{\bibinfo{journal}{\mnras}} \textbf{\bibinfo{volume}{349}},
  \bibinfo{pages}{735--741} (\bibinfo{year}{2004}).

\bibitem{motte07}
\bibinfo{author}{{Motte}, F.} \emph{et~al.}
\newblock \bibinfo{title}{{The earliest phases of high-mass star formation: a 3
  square degree millimeter continuum mapping of Cygnus X}}.
\newblock \emph{\bibinfo{journal}{\aap}} \textbf{\bibinfo{volume}{476}},
  \bibinfo{pages}{1243--1260} (\bibinfo{year}{2007}).

\bibitem{braine07}
\bibinfo{author}{{Braine}, J.}, \bibinfo{author}{{Ferguson}, A.~M.~N.},
  \bibinfo{author}{{Bertoldi}, F.} \& \bibinfo{author}{{Wilson}, C.~D.}
\newblock \bibinfo{title}{{The Detection of Molecular Gas in the Outskirts of
  NGC 6946}}.
\newblock \emph{\bibinfo{journal}{\apjl}} \textbf{\bibinfo{volume}{669}},
  \bibinfo{pages}{L73--L76} (\bibinfo{year}{2007}).

\bibitem{parravano03}
\bibinfo{author}{{Parravano}, A.}, \bibinfo{author}{{Hollenbach}, D.~J.} \&
  \bibinfo{author}{{McKee}, C.~F.}
\newblock \bibinfo{title}{{Time Dependence of the Ultraviolet Radiation Field
  in the Local Interstellar Medium}}.
\newblock \emph{\bibinfo{journal}{\apj}} \textbf{\bibinfo{volume}{584}},
  \bibinfo{pages}{797--817} (\bibinfo{year}{2003}).

\bibitem{matzner00}
\bibinfo{author}{{Matzner}, C.~D.} \& \bibinfo{author}{{McKee}, C.~F.}
\newblock \bibinfo{title}{{Efficiencies of Low-Mass Star and Star Cluster
  Formation}}.
\newblock \emph{\bibinfo{journal}{\apj}} \textbf{\bibinfo{volume}{545}},
  \bibinfo{pages}{364--378} (\bibinfo{year}{2000}).

\bibitem{huff06}
\bibinfo{author}{{Huff}, E.~M.} \& \bibinfo{author}{{Stahler}, S.~W.}
\newblock \bibinfo{title}{{Star Formation in Space and Time: The Orion Nebula
  Cluster}}.
\newblock \emph{\bibinfo{journal}{\apj}} \textbf{\bibinfo{volume}{644}},
  \bibinfo{pages}{355--363} (\bibinfo{year}{2006}).

\bibitem{feigelson07}
\bibinfo{author}{{Feigelson}, E.~D.} \& \bibinfo{author}{{Townsley}, L.~K.}
\newblock \bibinfo{title}{{The Diverse Stellar Populations of the W3 Star
  Forming Complex}}.
\newblock \emph{\bibinfo{journal}{\apj}}  (\bibinfo{year}{2007}).
\newblock \bibinfo{note}{In press, arXiv:0710.0090}.

\bibitem{mckee02}
\bibinfo{author}{{McKee}, C.~F.} \& \bibinfo{author}{{Tan}, J.~C.}
\newblock \bibinfo{title}{{Massive star formation in 100,000 years from
  turbulent and pressurized molecular clouds}}.
\newblock \emph{\bibinfo{journal}{\nat}} \textbf{\bibinfo{volume}{416}},
  \bibinfo{pages}{59--61} (\bibinfo{year}{2002}).

\end{thebibliography}
\end{document}